\documentclass[aps,prb,twocolumn,groupedaddress,showpacs]{revtex4}
\usepackage[latin1]{inputenc}        
\usepackage{isolatin1}
\usepackage{graphicx}
\usepackage{bm}
\usepackage{amssymb, amsmath}

\begin{document}

\title{Influence of Collision Cascade Statistics on Pattern
  Formation of Ion-Sputtered Surfaces}

\author{M.\ Feix}
\author{A.\ K.\ Hartmann}
\email[]{hartmann@theorie.physik.uni-goettingen.de}
\author{R.\ Kree}
\email[]{reiner.kree@physik.uni-goettingen.de}
\affiliation{Institut für Theoretische Physik, Universität Göttingen,
              Göttingen, Germany}
\author{J.\ Mu\~noz-Garc\'ia}
\email[]{jamunoz@math.uc3m.es}
\author{R.\ Cuerno}
\email[]{cuerno@math.uc3m.es}
\affiliation{Departamento de Matem\'aticas and Grupo Interdisciplinar de
              Sistemas Complejos (GISC), Universidad Carlos III de Madrid,
              Avenida de la Universidad, 30, 28911 Legan\'es, Spain}


\date{\today}

\begin{abstract}
Theoretical continuum models that describe the formation of patterns on
surfaces of targets undergoing ion-beam sputtering, are based on Sigmund's
formula, which describes the spatial distribution of the energy deposited by
the ion. For small angles of incidence and amorphous or polycrystalline
materials, this description seems to be suitable, and leads to the classic
BH morphological theory [R.\ M.\ Bradley and J.\ M.\ E.\
Harper, J.\ Vac.\ Sci.\ Technol.\ A {\bf 6}, 2390 (1988)].
Here we study the sputtering of Cu crystals by means of numerical simulations
under the binary-collision approximation. We observe significant deviations
from Sigmund's energy distribution. In particular, the distribution that best
fits our simulations has a minimum near the position where the ion penetrates
the surface, and the decay of energy deposition with distance to ion trajectory
is exponential rather than Gaussian. We provide a modified continuum theory
which takes these effects into account and explores the implications of the
modified energy distribution for the surface morphology.
In marked contrast with BH's theory,
the dependence of the sputtering yield with the angle of incidence is
non-monotonous, with a maximum for non-grazing incidence angles.
\end{abstract}


\maketitle

\section{\label{intro} Introduction}
Ion bombardment of solids often gives rise to characteristic surface
topographies, which evolve under stationary and homogeneous ion
fluxes. Besides kinetic roughening, wavelike ripple structures may
occur. Such height modulations on the submicron scale have been
observed for crystalline semiconductors \cite{lewis:1980,chason:1994}
as well as for crystalline metals \cite{rusponi:1998b,habenicht:1999}
and some amorphous \cite{mayer:1994} and polycrystalline materials, see a
recent review in Ref.\ \onlinecite{valbusa:2002}. According to continuum
theories, which are based on the work of
Bradley and Harper (BH),\cite{bradley_harper:1988}
the periodic patterns emerge from a competition between a roughening curvature
instability due to characteristics of the spreading of ion energy, and
simultaneous smoothing processes due to surface
diffusion.\cite{cuerno:1995,makeev:2002} Although this mechanism
seems to be quite universal, there
are material-specific differences in the evolution of surface
topographies. For non-metallic substrates, for example, one usually
needs off-normal incidence of ion flux to produce ripples, which change their
orientation with the incidence angle,\cite{mayer:1994,chason:1994,demanet:1995,maclaren:1992, carter:1996,malherbe:1994,erlebacher:1999,habenicht:1999}
whereas ripples are observed on metallic substrates even at normal
incidence, and the orientation of ripples changes with substrate
temperature.\cite{rusponi:1997,rusponi:1998,rusponi:1998b}
Furthermore, the smoothing mechanism of surface diffusion is not
well understood yet. In previous simulations,\cite{hartmann:2002} we have found
that the emerging patterns depend crucially on the diffusion mechanisms applied.
In particular the long-time behavior, which is governed in the continuum theory
by non-linear terms, depends even qualitatively on the surface diffusion mechanism.
Given that the surface topographies resulting from different mechanisms of
surface diffusion have been studied by simulations elsewhere,\cite{yewande:2004}
in the present work we will focus on specificities due to the energy deposition
process.

Continuum theories for the surface morphology of the target
usually assume that the kinetic energy of an ion hitting a solid surface
spreads in the bulk and produces a Gaussian density of deposited energy
\begin{align}
    \epsilon_s({\bf r} )&= N_s\,\epsilon\,
     e^{-\frac{x^2+y^2}{2\beta^2}}e^{-\frac{(z+a)^2}{2\alpha^2}},
    \label{distrib1}\\
    N_s&=\big[(2\pi)^{3/2}\alpha\beta^2\big]^{-1},\nonumber
\end{align}
where ${\bf r}=(x,y,z)$ is a point within the target, ions are falling along
the $\mathbf{\hat{z}}$ axis and penetrate an average distance $a$ within
the solid, $\epsilon$ is the average kinetic energy carried by each ion, and
the values of $\alpha$, $\beta$ describe the spreading of the energy,
they are of the same order of magnitude as $a$.
The Gaussian form (\ref{distrib1}) is based on the work of
Sigmund,\cite{sigmund:1969}
who considered a polycrystalline or amorphous target and analyzed the
kinetic transport theory of the sputtering process. He found that in
the elastic collision regime at energies where electronic stopping is
not dominating, the deposited energy can be approximated by a Gaussian
near its maximum. The quality of the approximation is reasonable, if
mass differences between substrate and ion are not too
large. Obviously, the Gaussian form is not universal and consequences
of deviations from the Gaussian form within the BH model
have not been studied yet. In particular, although the observations of
ripples on single crystalline
metals\cite{rusponi:1997,rusponi:1998,rusponi:1998b} are
qualitatively described by the BH model, the latter is strictly a theory
for amorphous materials, and thus there is a need to justify theoretically
the emergence of such type of patterns onto this other class of substrates.

Obtaining more detailed information about the deposited energy from
simulations has become straightforward by now, as there are many well
calibrated, efficient simulation methods for ion impact
available.\cite{trim, tombrello:1993,robinson:1974,koponen:1997, robinson:1994}
In the present work, we use simulations based on the binary collision
approximation\cite{robinson:1974,robinson:1994, koponen:1996} and
consider a metallic material (Cu), for which we
generate statistical ensembles of collision cascades emerging from
single ion impact events on plane
surfaces. We analyze the data in terms of deposited energies as well
as ejected particles. We do not claim to perform the best state of the
art simulation of ion impact on Cu (for example, we only consider
a very rough model of surface binding forces). Rather, we aim at
more generic results, which are of relevance to the theory of surface
evolution. Our simulations provide an average density of deposited
energy, which 
is quite different from the Gaussian form in Eq.\ (\ref{distrib1}).
We furthermore consider the fluctuations around this average and find
strong, intrinsic noise. In the subsequent part of our work we
investigate the consequences of the simulation results for the
continuum theory of pattern formation by ion-beam sputtering. 
We obtain that the modified energy
distribution obtained in the numerical simulations induces a sputtering yield
that overcomes some of the shortcomings (when comparing with experiments)
of the analogous result within BH's theory. Moreover, we recover the production
of the ripple instability, and the dependence of the pattern features with
phenomenological parameters similar to BH theory, thus providing a theoretical
framework within which observations of ripples on metals can be naturally
accomodated.\cite{rusponi:1997,rusponi:1998,rusponi:1998b}

\section{\label{method} Observables of Cascade Statistics}

In this section we want to relate
observables of our simulations to the phase space density
\begin{equation}
g(\bm{v},\bm{\rho},z,t|\bm{\rho}_0,z=0,\bm{v}_0,t=0) ,
\end{equation}
where $\bm{\rho} \equiv (x,y)$. This function
is the basic quantity underlying the kinetic
theory of collision cascades and also
introduces the quantities which are used in the construction of a continuum
theory of surface pattern formation by ion bombardment. Function
$g$ is the average density of cascade particles in 6-dimensional
$(\bm{v},\bm{\rho},z)$ space at time $t$,
under the assumption that one ion has hit the surface at $\bm{\rho}_0$ and
at $t=0$ with velocity $\bm{v}_0$. As we will only treat identical
initial conditions with $\bm{\rho}_0=0$ and
$\bm{v}_0=-|\sqrt{2\epsilon_0/m}|\bm{e}_z$, we
will use the abbreviated notation $g(\bm{\rho},z,\bm{v},t)$ and drop the
explicit dependence on $\bm{\rho}_0$ and $\bm{v}_0$.
The average has to be taken over an ensemble of targets, which differ
by random, thermal displacements of atoms.

To define our simulation observables in terms of the phase space
density, first note that
$
g(\bm{\rho},z,\bm{v},t)(\bm{v}\cdot d\bm{a})~d^3\bm{v}dt
$
is the number of
particles, which  penetrate a surface element $d\bm{a}$ situated at
position $(\bm{\rho},z)$,  with velocity $\bm{v}$ during the time
interval $dt$.

The phase space density and the corresponding current density may as
well be considered as functions of
position, energy and direction of velocity using
$\bm{v}=\sqrt{2\epsilon/m}\hat{\bm{v}}$, $\hat{\bm{v}}$ being the unit
vector in the direction of $\bm{v}$, so that
$\bm{v}d^3\bm{v}=(2/m^2)\epsilon\hat{\bm{v}}~ d\epsilon d\hat{\bm{v}}$.
The current density $\bm{j}(\epsilon, \bm{\rho},z,t) d\epsilon$ of cascade
particles of energy near $\epsilon$ is given by
\begin{equation}
\label{eq:current_density}
\bm{j}(\epsilon,\bm{\rho},z,t)d\epsilon =\frac{2}
{m^2}d\epsilon\int d\hat{\bm{v}}~
\epsilon\hat{\bm{v}}g(\epsilon,\hat{\bm{v}},\bm{\rho},z,t).
\end{equation}
If we restrict the integration over $\hat{\bm{v}}$ to directions with
$\hat{\bm{v}}\cdot d\bm{a} > 0$, we only count particles, which cross
the surface element in the direction of its normal. This variant will
be denoted by $\bm{j}^+$.

From Eq.\ (\ref{eq:current_density}) it is obvious that the time integral
\begin{equation}
h_{2d}(\epsilon,\bm{\rho},z)~dxdy=\int_0^{\infty} dt~ \bm{e}_z\cdot
\bm{j}(\epsilon, \bm{\rho},z,t) ~dxdy
\end{equation}
equals the total average number of particles per energy  at energy
$\epsilon$ of a single collision
cascade, which penetrate the surface element $dxdy\bm{e}_z$ located at
$(\bm{\rho},z)$.

Note that $h_{2d}$ is a surface density and the quantities
\begin{equation}
\label{eq:n2d}
n_{2d}(\bm{\rho},z)=\int_{0}^{\infty} d\epsilon~ h_{2d}(\epsilon, \bm{\rho},z)
\end{equation}
and
\begin{equation}
\label{eq:e2d}
e_{2d}(\bm{\rho},z)=\int_{0}^{\infty} d\epsilon~
\epsilon h_{2d}(\epsilon, \bm{\rho},z)
\end{equation}
give the average number of particles per area and average energy per
area transported to the $(xy)$-plane at $z$ by the
collision cascade. For all these quantities the corresponding variants
$h_{2d}^+, n_{2d}^+, e_{2d}^+$ only take into account particles moving
in outward direction $+\bm{e}_z$.

The particles arriving at the $z=0$ plane (which constitutes the surface of the
material) with velocities in outward direction will leave the bulk if
they overcome the surface binding forces.
We will use a simple spherical barrier model of surface binding with
barrier height $U$. This implies that
all particles arriving at the surface with kinetic energy $\epsilon>U$
will be sputtered off. The surface density $n_U$ of these particles is
therefore given by Eq.\ (\ref{eq:n2d}) with the lower boundary of the
$\epsilon$ integration replaced by $U$.
The total sputtering yield is the surface integral of this density, $Y_U=\int d^2\bm{\rho}~
n_U(\bm{\rho})$.
At internal surfaces there is no surface binding and thus
\begin{equation}
p(\epsilon,\bm{\rho},z)= h_{2d}(\epsilon,\bm{\rho},z)/Y_{U=0}
\end{equation}
becomes the
probability density to find a particle with energy $\epsilon$ crossing the
internal surface at location $(\bm{\rho},z)$
and
$
p_U(\bm{\rho})=n_U(\bm{\rho})/Y_U$
is the
probability density to find a particle leaving the bulk at
$\bm{\rho}$. These are the quantities we will study in the
subsequently described simulations.


\section{\label{sec:bca} BCA Simulations}
\begin{figure}[!htbp]
\includegraphics[width=0.45\textwidth]{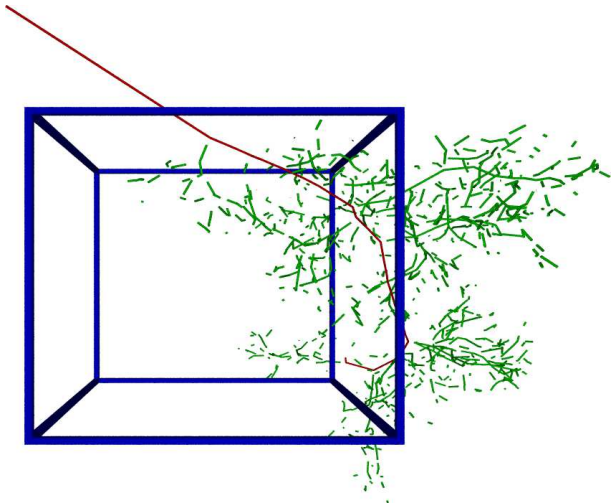}%
\caption{\label{fig:cascade} Sample cascade originating from an impact
of a 5 keV Cu ion on a Cu crystal. The angle of incidence is
$60^\circ$. The cube shown acts just as scale and 
has size $2.65$ nm$^3$, while the full
lattice simulated has size $(10.6$ nm$)^2\times 18$ nm.}
\end{figure}

Atomic displacements and particle ejection from a solid due to the impact of a
single ion with kinetic energy in the keV range can be simulated by using the
binary collision approximation \cite{robinson:1994} (BCA).
The basic idea is to substitute the detailed particle trajectories by
trajectories where the particles travel with constant velocity until
they ``hit'' onto another particle. Each collision event is integrated
analytically or numerically, leading to new positions and velocities
of the particles participating in the collision. Hence the full
dynamical process is reduced to a cascade of collisions. A sample cascade,
originating from an impact of a 5 keV Cu ion on a Cu lattice with an
angle of incidence of $60^\circ$, is shown in Fig.\ \ref{fig:cascade}.
Although this method has its
limitations,\cite{robinson:1994} it has become a standard technique and
is used to describe ion implantation and sputtering.

We have performed BCA simulations of single ion impact on a plane
Cu-surface at normal
incidence with velocity $\bm{v}_0=-|\bm{v}_{0}|\bm{e}_z$.

All the statistical information was obtained from ensembles of
3000-6000 ion impacts per ensemble, which we generated for a single
initial condition of the ion.  The positions of the atoms making up
the undisturbed solid were displaced from ideal lattice sites of a
Cu single  crystal (170000 atoms of an fcc structure with lattice
constant $3.61$ {\AA} corresponding to a  solid of $10.6 \; {\rm nm}
\times 10.6$ nm and a bulk depth of 18 nm) by uncorrelated,  Gaussian
distributed displacements to account for thermal fluctuations. For
each ion of an ensemble, an additional homogeneous lateral random
displacement was added, which was taken to be uniformly  distributed
within a square of edge length 1 lattice constant. Thus within every
ensemble the ion hits  upon macroscopically identical but
microscopically differing configurations of the solid.

We considered two orientations of the crystal,  $(1,0,0)$ and  $(58,
72, \overline{39})$. The latter was used to suppress effects of crystal
anisotropy and we found very good agreement between the angular
averages of $n_{2d}, ~e_{2d}$ obtained from $(1,0,0)$ and the
corresponding quantities obtained from the oblique orientation.

Our BCA code follows standard implementations.\cite{robinson:1974} It
allows for arbitrary  positions of the bulk atoms and is suitable for
studying defect accumulation during multiple impact  (although this
feature is not used in the present work). Simple and well-tested forms
of the screened Coulomb potential \cite{moliere:1947} and  the
inelastic processes \cite{robinson:1974,firsov:1959} have been
chosen. All model parameters of the algorithm have been adjusted to
Cu projectiles of a few keV, hitting a Cu single crystal.\cite{footnote1}
These choices allowed for easy calibration and comparisons of
our implementation against literature results.  It should be
emphasized, however, that the main focus of the present work is on
generic results, which are of relevance for pattern formation of
ion-sputtered surfaces of metals.

\section{\label{sec:pos_energy} Simulation Results}

\begin{figure}[!htbp]
\includegraphics[width=0.45\textwidth]{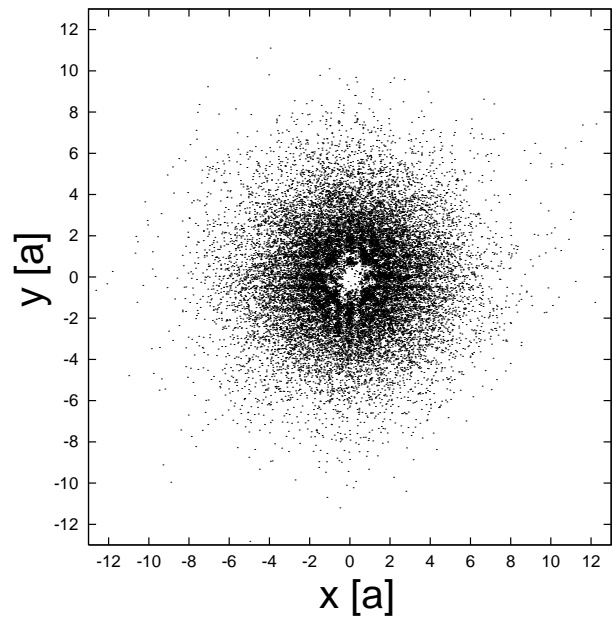}%
\caption{\label{fig:cloud100} Spatial distribution of ejected Cu atoms
  emerging from 6000 independent trials of hitting the $(x,y)$ crystal
  surface (oriented in  $(1,0, 0)$ direction) with a
  single 5 keV Cu ion at normal incidence. Distances are measured in
  units of $a=3.61$ \AA.  }
\end{figure}

\begin{figure}[!htbp]
\includegraphics[width=0.45\textwidth]{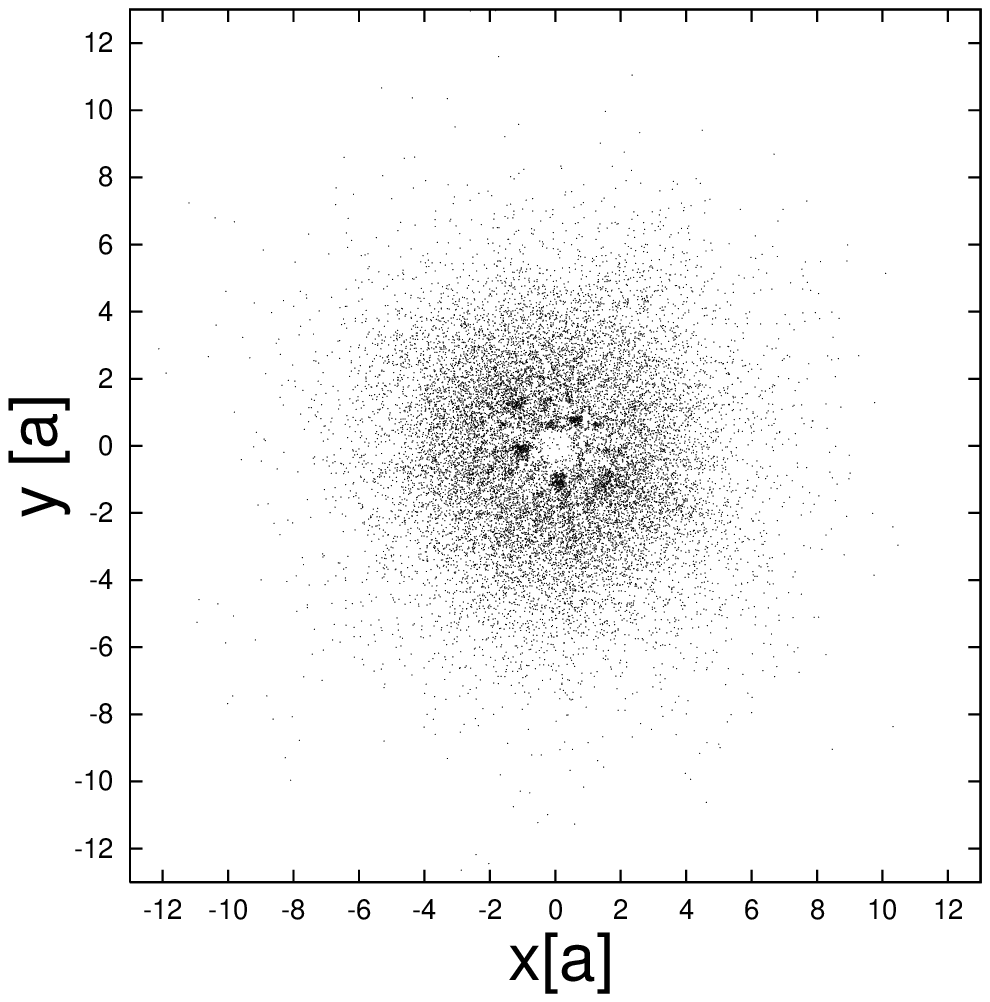}%
\caption{\label{fig:cloud} Spatial distribution of ejected Cu atoms
  emerging from 6000 independent trials of hitting the $(x,y)$ crystal
  surface (oriented in  $(58, 32, \overline{39})$ direction) with a
  single 5 keV Cu ion at normal incidence. Distances are measured in
  units of $a=3.61$ \AA.  }
\end{figure}

Fig.\ \ref{fig:cloud100} shows the surface distribution of all the ejected
particles within an ensemble of  6000 cascades, each  emerging from
one incident 5 keV Cu ion (normal incidence) for the crystal in
(1,0,0) orientation. Clearly, a  ``hole''
around the location of impact is visible. This is in contrast to what
can be expected when applying Eq.\ (\ref{distrib1}).
The scattering is almost rotational invariant, a slight $90^\circ$
rational-invariant structure is visible, reflecting the lattice structure.
To check whether the result is an artefact of the crystal orientation,
we studied also a $(58,32,\overline{39})$ surface, see Fig.\ \ref{fig:cloud}.
Although the plot exhibits slightly less structures, again only few
particles are ejected near the point of penetration.
Hence, we decided to concentrate on the oblique $(58, 32,
\overline{39})$ orientation, because
we want to study generic results irrespective of specific
crystal orientations.
In any case, for both surface orientations and small
values of $\Delta$, the angular average of data obtained from
\begin{equation}
  \label{eq:angularaverage}
n_{2d}(\rho)=\frac{1}{2\pi\rho\Delta}~\int_0^{2\pi}d\phi
\int_{\rho}^{\rho+\Delta}\rho \, d\rho ~n_{2d}(\rho,\phi) ,
\end{equation}
where $\rho = |{\bm{\rho}}|$, turns out to be nearly undistinguishable.

\begin{figure}[!htbp]
\includegraphics[width=0.5\textwidth]{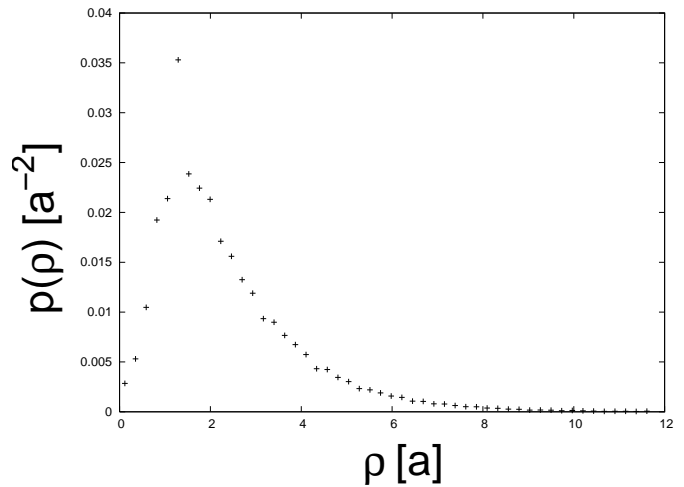}%
\caption{\label{fig:p_r} Probability of ejected particles vs distance
  $\rho$  from point of ion
  incidence (measured in  units of $a=3.61$ \AA)
determined from the data of  Fig.\ (\ref{fig:cloud}). }
\end{figure}

Fig.\ \ref{fig:p_r} shows the corresponding probability density
$p(\rho)$ per surface unit, averaged over all angles,
of finding an ejected particle a distance $\rho$ from the
point of incidence of the Cu ion. This figure shows that the
assumption on the ejection probability being distributed following
a Gaussian distribution, hence leading to a maximum at $\rho=0$, is not
justified in case of crystals. This is in contrast to amorphous materials,
where a more Gaussian-like distribution of the ejection probability has been
observed \cite{hofsaess:2004} when using the simulation packet SRIM with the
same ion/bulk parameters as above.

\begin{figure}[!htbp]
\includegraphics[width=0.5\textwidth]{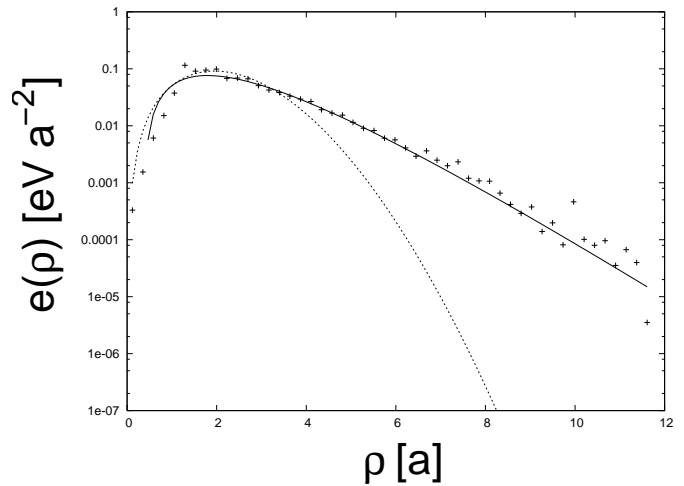}%
\caption{\label{fig:e_r} Surface density of mean energy of sputtered Cu
atoms vs distance $\rho$ (measured in units of $a=3.61$ \AA)
from point of ion incidence for 5 keV Cu ions on semi-log scale.
The solid line is the best fit of the data to an exponential with a
polynomial prefactor and corresponds to
$0.297(\rho^2-0.392\rho)\exp(-1.27\rho)$. The dotted line,
which corresponds to a fit to a Gaussian, is obviously inadequate.}
\end{figure}

Fig.\ \ref{fig:e_r} displays the corresponding angular average of the
surface density $e_{2d}(\rho)$ of the energy of sputtered particles.
In this figure, we have also shown two
Marquardt-Levenberg fits ($f_s=(a\rho^2+b\rho)\exp[-c\rho^s]$ with $s=1$
and $s=2$) to the data. One can see that the decay of the energy
density is not in accordance with a Gaussian, even when including a
decay towards the point $\rho=0$ of penetration, as suggested by
Eq.\ (\ref{distrib1}).  The data can be fitted well to a exponential
decay with a simple polynomial prefactor.

\begin{figure}[!htbp]
\includegraphics[width=0.5\textwidth]{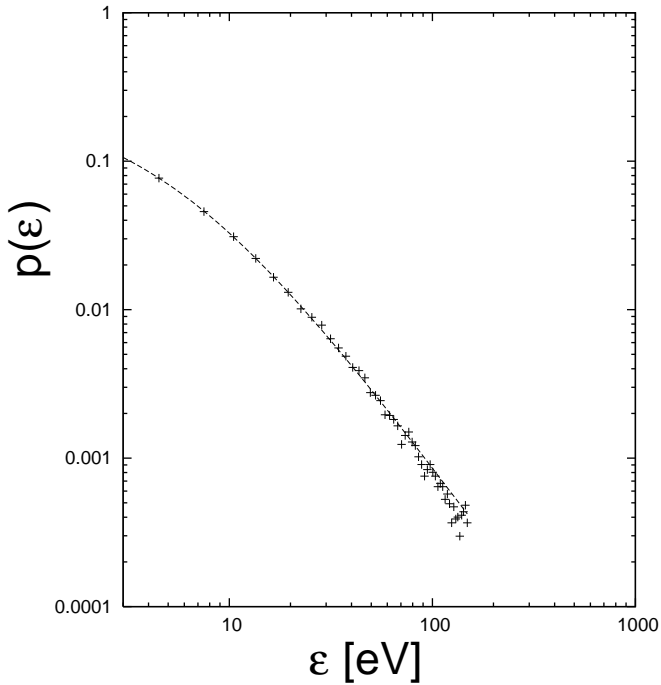}%
\caption{\label{fig:p_e} Probability density of energy, which is
  transported to the surface by a single collision cascade emerging
  from a 5 keV Cu ion on a log-log scale.
The solid line corresponds to $5.259(5.035+\epsilon)^{-1.874}$, which is the
best fit to a simple power law $a/(b+\epsilon)^\alpha$.}
\end{figure}

In the previous figure, we have studied the mean, hence
let us now consider characteristic features of the probability density
$p(\rho,\epsilon)$.
In Fig.\ \ref{fig:p_e} we show the surface-integrated probability density
$p(\epsilon)=\int d^2\bm{\rho}~ p(\epsilon,\rho)$. The behavior is
remarkable close to a simple power law
\begin{equation}
p(\epsilon)\approx \frac{a}{(b+\epsilon)^\alpha} \sim \epsilon^{-2}
\end{equation}
outside of a region of small $\epsilon$. Our best fit corresponds to
$a=5.26$, $b=5.03$, $\alpha=1.87$.

\begin{figure}[!htbp]
\includegraphics[width=0.5\textwidth]{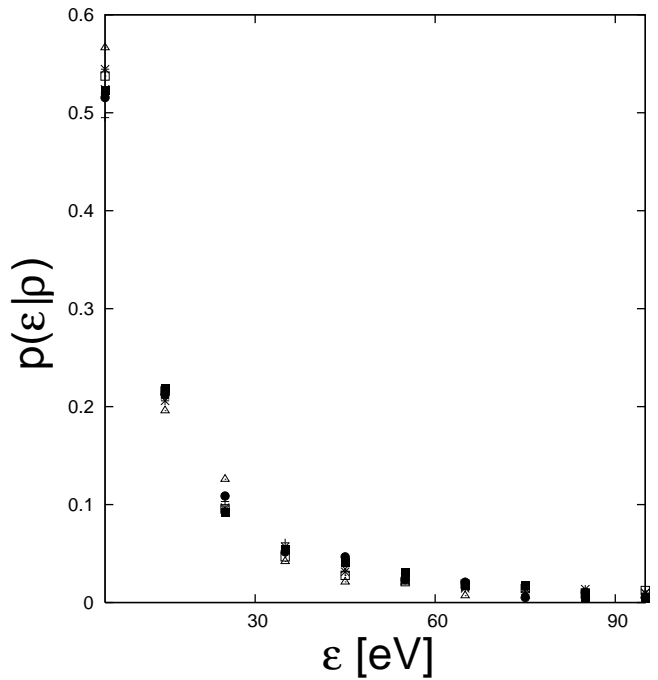}%
\caption{\label{fig:cond_prob} Conditional probability density
  $p(\epsilon|\rho)$ to find energy $\epsilon$ of ejected particles
  keeping the distance $\rho$ fixed. Different symbols correspond to
  different values $\rho= 1.759, 2.914, 4.104, 5.277, 6.450, 7.623,
  8.795, 9.968, 11.141$ As the data for different $\rho$ are almost
  undistinguishable, differences not being significant, we need not provide
  a detailed legend.}
\end{figure}

Fig.\ \ref{fig:cond_prob} displays the conditional probability density
$p(\epsilon|\rho)$ of energy at fixed $\rho$ for different values of $\rho$.
Surprisingly, the conditional density does not
depend on $\rho$ significantly. This shows that
\begin{equation}
p(\epsilon,\rho)\approx
p(\rho)p(\epsilon)\propto\frac{an_{2d}(\rho)}{(b+\epsilon)^2}
\end{equation}
outside of a region of very small distances.
An immediate implication is that
\begin{equation}
e_{2d}(\rho) \propto n_{2d}(\rho)
\end{equation}
so that the number of ejected particles and the energy deposited at
the surface are proportional to each other, as assumed in the BH
theory. However, another important
implication is that the amount of energy transported to the surface is
subjected to strong internal noise, which may limit the applicability
of deterministic continuum theories based on the average energy at the
surface. This problem will be pursued elsewhere.

\begin{figure}[!htbp]
\includegraphics[width=0.5\textwidth]{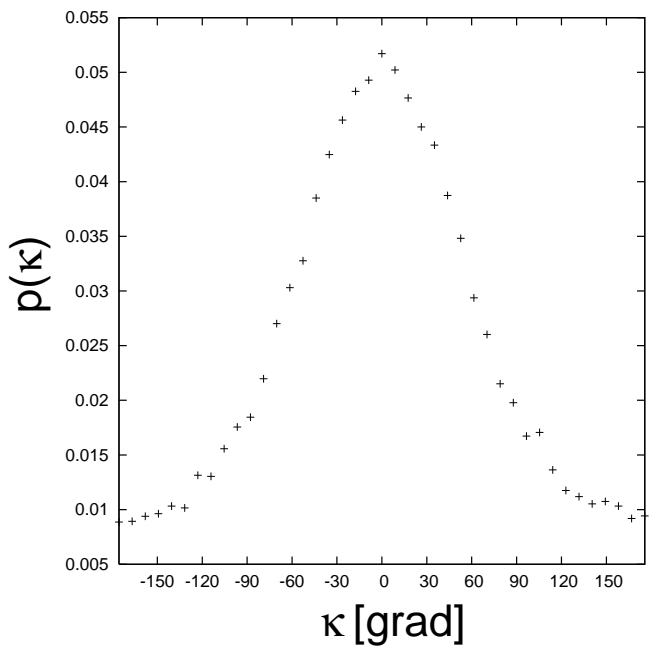}%
\caption{\label{fig:angle} Distribution of ejected particles with
  different directions of velocity. Here $\kappa$ denotes the angle between
  the projection of particle velocity onto the surface and the vector
  between point of ion impact and point of particle ejection.}
\end{figure}

To gain further insight into the structure of the ensemble of cascade
particles at the surface, we have also studied correlations between
the position $\bm{\rho}$ and the projection of velocity,
of sputtered particles onto the surface,
$\bm{v}_{surf}=\bm{v}-(\bm{v}\cdot\bm{e}_z)\bm{e}_z$ .
 Fig.\ \ref{fig:angle} shows the distribution of the
angle $\kappa$ between $\bm{\rho}-\bm{\rho}_0$ and $\bm{v}_{surf}$, where
$\bm{\rho}_0$ is the point at which the ion hits the surface and
$\bm{\rho}$ is the position of a cascade particle arriving at the
surface with velocity $\bm{v}$. The figures shows that,
for typical collision cascades, most
of the ejected particles move away from the point of ion impact.

\section{Continuum approximation to energy deposition}
\label{secdm}

Within Sigmund's approximation,\cite{sigmund:1973} the rate at which the
target is being eroded at an arbitrary point on the surface, is proportional
to the total amount of energy deposited there from ion collisions. In his
theory for amorphous or polycrystalline targets, an accurate description of
the sputtering phenomena can be achieved by assuming that energy is deposited
following the Gaussian distribution (\ref{distrib1}).

Bradley and Harper (BH)\cite{bradley_harper:1988} later employed this energy
distribution in order to compute the {\em local} erosion velocity at an
arbitrary surface point $O$, allowing for gentle surface undulations.
To perform the calculation, a new local reference frame is taken in which
the $\mathbf{\hat{z}'}$ axis is taken along the surface normal at $O$.
The principal curvatures are assumed along the $\mathbf{\hat{x}'}$ and
$\mathbf{\hat{y}'}$ axes, that are defined, respectively, as the direction
orthogonal to $\mathbf{\hat{z}'}$ that is in the plane defined by this
axis and the ion trajectory and the remaining direction in order to make up
a right-handed reference frame. Assuming that the radii of curvature
at $O$, $R_x$ and $R_y$,  are much larger than the penetration depth $a$,
the surface height can be approximated to
$z'(x',y')=-\frac{1}{2}(\frac{x'^2}{R_x}+\frac{y'^2}{R_y})$.
In order to obtain the erosion velocity, we have to add up the total
energy deposited at $O$ from ions entering the whole target, expressing
the ion flux and energy distribution in the latter reference frame, which
is related with the one implicit in (\ref{distrib1}) as
\begin{align}\label{eq:cylindrical}
\mathbf{\hat{x}}&=\mathbf{\hat{x}'}\, \cos(\gamma_0) + \mathbf{\hat{z}'} \,
\sin(\gamma_0),\nonumber\\
\mathbf{\hat{y}}&=\mathbf{\hat{y}'},\\
\mathbf{\hat{z}}&=\mathbf{\hat{z}'} \cos(\gamma_0) -
\mathbf{\hat{x}'}\,\sin(\gamma_0),\nonumber
\end{align}
with $\gamma_0$ being the incidence angle formed between the ion
trajectories and the surface normal at $O$. Accounting up to
curvature corrections, the ion flux reads $\Phi({x',y'})=\Phi_0
\cdot(\cos\gamma_0 - \frac{x'}{R_x}\sin\gamma_0)$, where $\Phi_0$ 
is the {\em constant} nominal ion flux. Taking all this into account, the erosion
velocity at $O$ reads, finally,
\begin{equation}\label{2}
v_O(\gamma_0 , R_x , R_y)=\Lambda
\int^{+\infty}_{-\infty}\int^{+\infty}_{-\infty}\Phi({x',y'})\,\epsilon_s(x',y')
\,dx'\,dy',
\end{equation}
where $\Lambda$ is a proportionality constant relating deposited energy with
the number of sputtered atoms, and the integration limits are taken to
infinity thanks to the fast decay of the energy distribution $\epsilon_s=e_{2d}$.
By expanding (\ref{2}) to lowest non-trivial order in $a/R_x$, $a/R_y \ll
1$, Bradley and Harper obtained\cite{bradley_harper:1988}
\begin{align}\label{eqvo}
v_O&= N_s\Lambda \epsilon\Phi_0 e^{-\frac{a^2}{2\alpha^2}}
\Big[\Gamma_0(\gamma_0)+
\frac{\Gamma_x(\gamma_0)}{R_x}+\frac{\Gamma_y(\gamma_0)}{R_y}\Big],
\end{align}
where $\Gamma_0(\gamma_0)$, $\Gamma_x(\gamma_0)$, and $\Gamma_y(\gamma_0)$
are functions that depend on the incidence angle $\gamma_0$, but also on
features of the energy distribution such as $a$, $\alpha$, and $\beta$.

Formula (\ref{eqvo}) enables computation of various relevant observables.
Thus, the sputtering yield, $Y(\gamma_0)$, defined as the total number
of sputtered atoms per incident ion, is easily related to $v_O$ by
geometry as $Y(\gamma_0)=n\,v_O(\gamma_0)/(\Phi_0\,\cos\gamma_0)$,
where $n$ is the number of atoms per unit volume in the target.
Assuming a planar interface, that is, in the $R_x, R_y
\rightarrow \infty$, one is left with
\begin{equation}
    Y(\gamma_0)=\frac{n\,v_O(\gamma_0,R_x \rightarrow \infty, R_y\rightarrow
    \infty)}{\Phi_0\,\cos\gamma_0}.
    \label{Y0}
\end{equation}
Working with Sigmund's distribution, BH found\cite{bradley_harper:1988} that
$Y(\gamma_0)$ increases monotonously as a function of the indigence angle
$\gamma_0$, such that the maximum efficiency for erosion is achieved
at grazing incidence, contrary to experimental evidence for amorphous,
polycrystalline, and crystalline
targets.\cite{Oechsner:1975,Cheney:1965,Olivia:1987} This feature of BH's
theory originates in a property of Sigmund's distribution (\ref{distrib1}),
whose maximum for deposition, ${\bf r}=(0,0,-a)$, is located right at the
surface under grazing incidence conditions. However, as is well known, there
usually exists a value of $\gamma_0 < 90^{\circ}$ for which the yield is
maximum, such that the sputtering efficiency decreases for larger angles of
incidence, due to ions being reflected at the surface, an effect which is
beyond Sigmund's approximations.

Additional predictions on the morphology of the eroded target can be
derived from (\ref{eqvo}). Thus, $\Gamma_x(\gamma_0)$ is
negative\cite{bradley_harper:1988} for small angles of incidence, which
implies that the erosion velocity is larger at troughs ($R_x<0$) than
at peaks ($R_x>0)$, inducing a morphological instability. Additional
surface relaxation mechanisms exist, such as surface diffusion, that
counteract this instability. Competition between the two opposing phenomena
induce the emergence of a typical length scale, associated with the wavelength
of the periodic ripple structure appearing
(for details, see Sec.\ \ref{sececcont}). For $\gamma_0$
large enough, BH get $\Gamma_x>0$, while $\Gamma_y$ is always negative.
Given that at small angles $\Gamma_x<\Gamma_y$, one obtains that the
ripple crests are oriented perpendicular to the $\mathbf{\hat{x}'}$
direction for incidences close to normal, whereas they are
oriented perpendicular to the $\mathbf{ \hat{y}'}$ direction
for incidence angles larger than a critical one, $\gamma_0 > \gamma_c$
such that $\Gamma_y<\Gamma_x$. Many experiments\cite{chason:1994,Vajo:1996}
have verified the validity of BH's theory to describe ripple wavelength and
orientation.

\section{Modified energy distribution functions}

The results of computer simulations within the BCA approximation,
obtained in the previous sections for Cu ion bombardment of a ${\rm Cu}$
target are described by an energy distribution that differs substantially
from that obtained by Sigmund in the case of polycrystalline or amorphous
substrates. Using cylindrical coordinates around the ion trajectory,
as in previous sections, we have [recall Fig.\ \ref{fig:e_r} and Eq.\ 
(\ref{eq:cylindrical}) above]
\begin{equation}
\epsilon_e(\bm{\rho},z)=N_e\,
\epsilon \,(\rho^2+c\rho)\,e^{-\frac{\rho}{\sigma_{xy}}}\,
e^{-\frac{(z+a)^2}{2\sigma_z^2}},
\label{distrib2}
\end{equation}
where $N_e=[(2\pi)^{3/2}\sigma_z(6\sigma_{xy}^4+2c\sigma_{xy}^3)]^{-1}$
is a normalization constant.
Values for $c$ and $\sigma_{xy}$ that best fit simulation results
were $c=-0.392$, $\sigma_{xy}=0.787$, see Fig.\ \ref{fig:e_r}.
Note two main differences of distribution
(\ref{distrib2}) to Sigmund's distribution (\ref{distrib1}): decay here is
slower [exponential as compared to Gaussian, thus the subscripts in
(\ref{distrib2})] in the plane perpendicular to the ion trajectory, and energy
deposition is {\em null} along the ion trajectory itself. On the other hand,
distribution (\ref{distrib2}) is unphysical. since $c<0$ leads to
 {\em negative} probabilities for
small $\rho$ values. For this reason, and in order to facilitate
analytical results [unavailable for (\ref{distrib2}) in the physical case
of a two dimensional target, see below], we will also consider
the following modified Gaussian (thus the subscripts) distribution
\begin{align}
\epsilon_g(\bm{\rho},z)&=N_g\,
\epsilon \,\rho^2\,e^{-\frac{\rho^2}{2\sigma_{xy}^2}}\,e^{-\frac{(z+a)^2}{2\sigma_z^2}},\label{distrib3}\\
N_g&=\big[2(2\pi)^{3/2}\sigma_{xy}^4\sigma_z\big]^{-1},\nonumber
\end{align}
that shares with (\ref{distrib2}) inducing zero energy deposition along
the ion trajectory $\rho=0$, but is otherwise Gaussian in all three
directions far enough from the ion path.

\subsection{One-dimensional interfaces}

In order to develop intuition about morphological predictions from
(\ref{distrib2}), (\ref{distrib3}), that are based on analytical
results, we consider first the (non-physical) case of a one-dimensional target, 
whose surface height is described by a single variable function $z'(x')$. 
These results will be then compared to the analogous
ones by Bradley and Harper, which will allow us to assess
differences due to the new form of the energy distribution, mostly
to the fact that in our case no energy is deposited along ion
trajectories.

For a one-dimensional target, distribution (\ref{distrib3}) reads
\begin{align}\label{distrib31d}
\epsilon^{1d}_g(x,z)&=N_g\,
\epsilon \,x^2\,e^{-\frac{x^2}{2\sigma_{x}^2}}\,e^{-\frac{(z+a)^2}{2\sigma_z^2}},\\
N^{1d}_g&=(2\pi\sigma_z\sigma_x^3)^{-1}.\nonumber
\end{align}
Writing the local velocity of erosion in terms of $\Gamma^{g,1d}_0$,
$\Gamma^{g,1d}_x$ analogous to Eq.\ (\ref{eqvo}), we obtain
\begin{equation}
v_O=N_g^{1d}\Lambda \epsilon \Phi_0\,e^{-\frac{a^2}{2\sigma_z^2}}\Big[\Gamma^{g,1d}_0
+ \frac{\Gamma^{g,1d}_x}{R_x}\Big],
\label{v1g}
\end{equation}
where the full expressions for $\Gamma^{g,1d}_0$ and $\Gamma^{g,1d}_x$ as
functions of $\gamma_0$, $a$, $\sigma_x$, and $\sigma_z$ can be found
in Appendix \ref{apendiceA}.

\begin{figure}[!htbp]
\begin{center}
\includegraphics[width=0.4\textwidth]{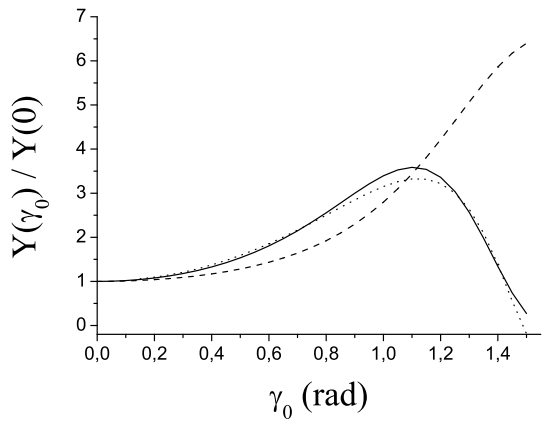}
\caption{\label{fig:Y01d} Normalized sputtering yield $Y(\gamma_0)/Y(0)$
as a function of incidence angle $\gamma_0$, for the various
one-dimensional energy distributions.
Dashed line: Bradley-Harper for $a=3.8$ nm, $\alpha=2.2$ nm, $\beta=1.5$ nm.
Solid line: modified Gaussian, Eq.\ (\ref{distrib31d}), for $a=3.8$ nm, $\sigma_z=2.2$
nm, $\sigma_x=1.5$ nm. Dotted line: exponential, Eq.\ (\ref{distrib21d}), for 
$a=3.8$ nm, $\sigma_z=2.2$ nm, $\sigma_x=0.787$ nm, $c=-0.392$ nm.}
\end{center}
\end{figure}

On the other hand, distribution (\ref{distrib2}) reads, for a one-dimensional
interface,
\begin{align} \label{distrib21d}
\epsilon^{1d}_e(x,z)&=N_e\,
\epsilon \,(x^2+c\,|x|)\,e^{-\frac{|x|}{\sigma_{x}}}\,e^{-\frac{(z+a)^2}{2\sigma_z^2}},\\
N^{1d}_e&=\big[\sqrt{2\pi}\sigma_z(4\sigma_x^3+2c\sigma_x^2)\big]^{-1}\nonumber.
\end{align}
In this case, the prediction for the local velocity of erosion has a shape that is
similar to (\ref{distrib31d}), albeit with more complex coefficients, whose
detailed analytical expressions are again left to Appendix \ref{apendiceB}:
\begin{equation}
    v_O=N^{1d}_e\Lambda \epsilon \Phi_0\,e^{-\frac{a^2}{2\sigma_z^2}}
    \Big[\Gamma^{e,1d}_0 + \frac{\Gamma^{e,1d}_x}{R_x}\Big].
    \label{v1e}
\end{equation}
In Fig.\ \ref{fig:Y01d}, we plot the normalized (to the corresponding values
for normal incidence) sputtering yields $Y(\gamma_0)$ obtained through
(\ref{Y0}) for the modified Gaussian (\ref{v1g}) and exponential (\ref{v1e})
distributions. For the sake of reference, the BH yield is also shown.
We can see that for both modified distributions, (\ref{distrib31d}) and
(\ref{distrib21d}), the corresponding yields feature maxima before
grazing incidence, as a difference to the BH curve. This is in agreement with
experimental data,\cite{Oechsner:1975,Cheney:1965,Olivia:1987} and is due
to the fact that maxima of energy deposition are not along the ion
trajectory for these distributions but, rather, at a certain finite distance
from it, what makes grazing incidence {\em not} be the most efficient for
sputtering. The fact (seen in Fig.\ \ref{fig:Y01d}) that the yield is
{\em negative} for large incidence angles $\gamma_0$, as computed
using the exponential distribution, is due to Eq.\ (\ref{distrib21d})
taking negative values for small distances to the ion path. As will be seen
below, this is an artifact of the one-dimensional approximation, as is
the fact that the yield computed from the modified Gaussian distribution
(\ref{distrib31d}) vanishes for $\gamma_0 = 90^{\circ}$.
From the figure we can also conclude that {\em qualitative} behaviors of
distributions (\ref{distrib31d}) and (\ref{distrib21d}) are similar, the
advantage of the first one being its greater analytical simplicity.
Parameters employed in Fig.\ \ref{fig:Y01d} are typical for
Cu ion bombardment of ${\rm Cu}$ for energies in the range of a few keV, as 
confirmed by TRIM/SRIM simulations.\cite{trim}

\subsection{Two-dimensional interfaces}

Naturally, the physically relevant case is bombardment of
two-dimensional targets. In this case, the analysis is more complex,
to the extreme that no closed expressions analogous of those found
previously for the exponential distribution (\ref{distrib2}) that
best fits our BCA simulation data. Results for this distribution
will be provided from numerical solutions of (\ref{2}) using
(\ref{distrib2}). On the other hand, we have seen in the 1d case
that distributions (\ref{distrib31d}) and (\ref{distrib21d}) lead
to similar qualitative results. For the 2d case, expression
(\ref{2}) using (\ref{distrib3}) leads to closed analytical
expressions for the coefficients in
\begin{equation} \label{vOg2d}
    v_O=N_g\Lambda \epsilon \Phi_0\,e^{-\frac{a^2}{2\sigma_z^2}}\Big[\Gamma^g_0 +
    \frac{\Gamma^g_x}{R_x}+\frac{\Gamma^g_y}{R_y}\Big],
\end{equation}
that can be found in Appendix \ref{apendiceC}.
In Fig.\ \ref{fig:Y02d} we see again that the sputtering yield for
both modified distributions (\ref{distrib2}) and (\ref{distrib3})
have maxima for incidence angles clearly smaller than grazing.
Moreover, the yields are positive and non-zero for all values of $\gamma_0$,
and amount to large sputtering rates, as found in
experiments.\cite{Oechsner:1975,Cheney:1965,Olivia:1987}.
In the present two-dimensional case, for grazing incidence
the radial component of the energy distribution vanishes at the point of impact
with the surface, but not at finite distances from it,
which implies that after surface integration the total deposited energy is
non-zero and the yield is positive. Again, Fig.\ \ref{fig:Y02d} shows
similar qualitative behaviors for both modified distributions, the
modified Gaussian having the advantage of leading to closed analytical
results.
\begin{figure}[!htbp]
\begin{center}
\includegraphics[width=0.4\textwidth]{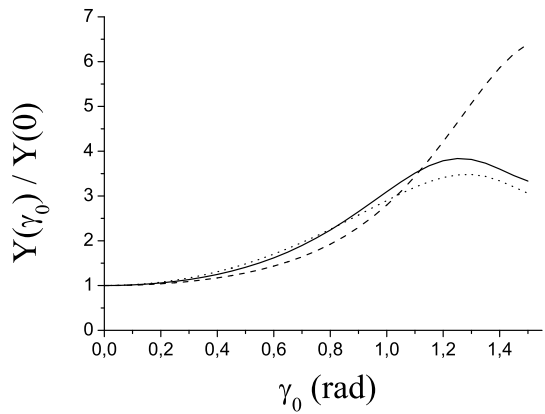}
\caption{\label{fig:Y02d} Normalized sputtering yield $Y(\gamma_0)/Y(0)$
as a function of incidence angle $\gamma_0$, for the various
two-dimensional energy distributions.
Dashed line: Bradley-Harper, Eq.\ (\ref{distrib1}), for $a=3.8$ nm,
$\alpha=2.2$ nm, $\beta=1.5$ nm. Solid line: modified Gaussian, Eq.\ (\ref{distrib3}),
for $a=3.8$ nm, $\sigma_z=2.2$ nm, $\sigma_{xy}=1.5$ nm. Dotted line: exponential, Eq.\ 
(\ref{distrib2}), for $a=3.8$ nm, $\sigma_z=2.2$ nm, $\sigma_x=0.787$ nm, $c=-0.392$ nm.}
\end{center}
\end{figure}
For normal incidence, the $x \leftrightarrow y$
symmetry is restored, and thus $\Gamma_x(\gamma_0)$ and $\Gamma_y(\gamma_0)$
must coincide. From Eq.\ (\ref{vOg2d}), we obtain
$\Gamma^g_x(0)=\Gamma^g_y(0)=-4\pi a \sigma_{xy}^6/\sigma_z^2$,
whose coincidence with the numerical results for distribution (\ref{distrib2})
has been confirmed. In Figs.\ \ref{fig:sGxy2d}, \ref{fig:gGxy2d}, and
\ref{fig:eGxy2d} we present results for the ``surface tension'' coefficients
$\Gamma_x$ and $\Gamma_y$ for the various two-dimensional distributions,
that are normalized by their corresponding absolute values for normal
incidence $\gamma_0=0$. We see in Figs.\ \ref{fig:gGxy2d}, \ref{fig:eGxy2d} that
$\Gamma_x$ is smaller than $\Gamma_y$ for incidence angles $\gamma_0
< \gamma_c$ and that $\Gamma_y$ is always negative, similarly to the
BH case (Fig.\ \ref{fig:sGxy2d}).
One of the successes of BH's theory lies in its description of
the orientation of the ripple structure for different ion incidence
angles $\gamma_0$. Here we see that, although distributions (\ref{distrib2}) and
(\ref{distrib3}), lead to quite different sputtering yields as compared to
Sigmund's distribution, the qualitative behavior of coefficients
$\Gamma_x$ and $\Gamma_y$ is quite similar to that found by BH.
Since experimental results are in good agreement with BH for metals,
the modified Gaussian distribution (\ref{distrib3}) seems a good choice for the
corresponding analytical description, with the advantage over the exponential 
distribution (\ref{distrib2}) of being physically sound for small distances to the 
point of penetration.
\cite{habenicht:1999,rusponi:1998b}
\begin{figure}[!htbp]
\begin{center}
\includegraphics[width=0.4\textwidth]{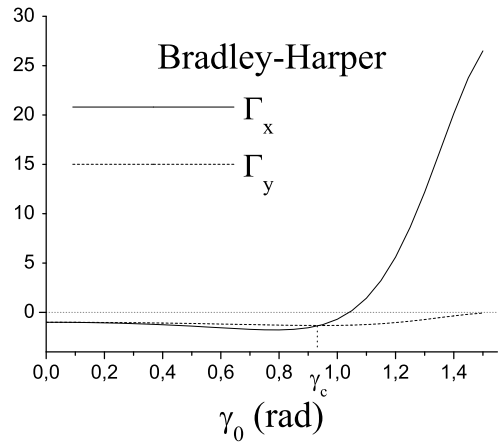}
\caption{\label{fig:sGxy2d} Normalized values of $\Gamma_x$ and $\Gamma_y$ 
for the distribution (\ref{distrib1}) using the same parameter values as in Fig.\ 
\ref{fig:Y02d}.}
\end{center}
\end{figure}
\begin{figure}[!htbp]
\begin{center}
\includegraphics[width=0.4\textwidth]{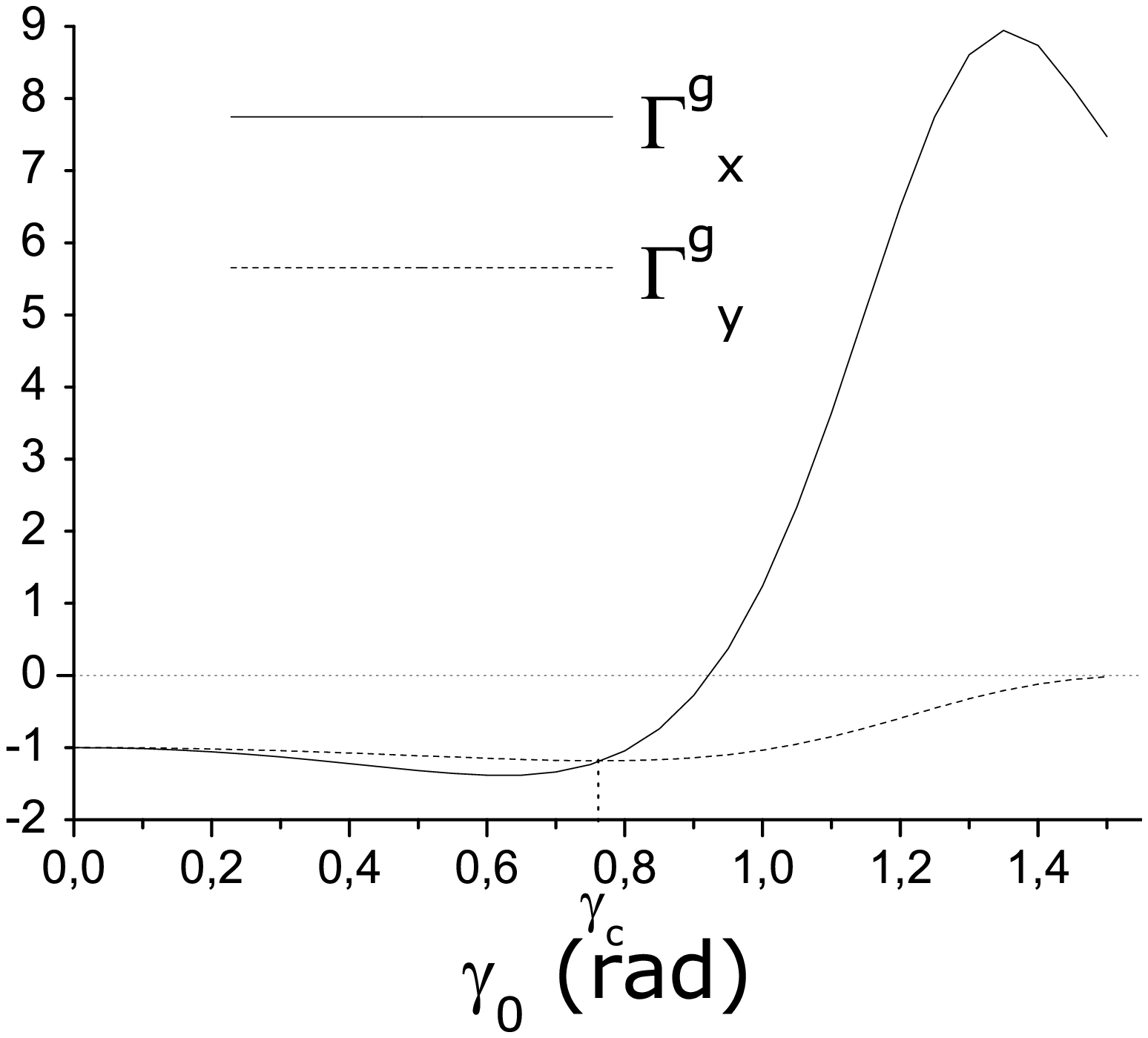}
\caption{\label{fig:gGxy2d} Normalized values of $\Gamma^g_x$ and $\Gamma^g_y$
for the distribution (\ref{distrib3}) using the same parameter values as in Fig.\
\ref{fig:Y02d}.}
\end{center}
\end{figure}
\begin{figure}[!htbp]
\begin{center}
\includegraphics[width=0.4\textwidth]{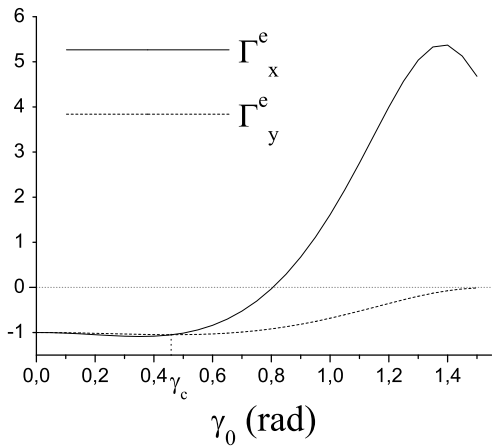}
\caption{\label{fig:eGxy2d} Normalized values of $\Gamma^e_x$ and $\Gamma^e_y$
for the distribution (\ref{distrib2}) using the same parameter values as in Fig.\
\ref{fig:Y02d}.}
\end{center}
\end{figure}

\section{Continuum equation for the surface height}
\label{sececcont}

Following the pioneering approach by Bradley and Harper, we can derive an
evolution (differential) equation for the surface height, starting from the
equation for the erosion velocity. We consider a laboratory frame
of reference ($\mathbf{\hat{X}}$, $\mathbf{\hat{Y}}$,
$\mathbf{\hat{Z}}$), defined as follows: the $\mathbf{\hat{Z}}$
axis is chosen to be normal to the initial planar surface. The
incoming beam direction forms an angle $\theta$ with
$\mathbf{\hat{Z}}$, and both direction define a plane where the
$\mathbf{\hat{X}}$ axis lies. Finally, the $\mathbf{\hat{Y}}$ axis
is perpendicular to the $\mathbf{\hat{X}}$ and $\mathbf{\hat{Z}}$
directions. We describe by $h(X,Y,t)$ the surface height at time $t$ above point 
$(X,Y)$ on reference plane of the unbombarded substrate, and assume that it varies 
slowly enough so we can work to first order in the derivates. In this way
we may approximate:\cite{bradley_harper:1988,cuerno:1995,makeev:2002} 
$\gamma_0=\theta-\frac{\partial h}{\partial
X}$, $\frac{1}{R_{x}}=-\frac{\partial^2 h}{\partial X^2}$,
$\frac{1}{R_{y}}=-\frac{\partial^2 h}{\partial Y^2}$. The velocity
of erosion of the surface height $h$ is provided by the
erosion rate $v_O$, and we thus get:
\begin{equation}\label{varh} \frac{1}{F} \frac{\partial h}{\partial t}\cong
-\Gamma_0(\theta)+\frac{\partial \Gamma_0(\theta)}{\partial
\theta} \frac{\partial h}{\partial X}+\Gamma_x \frac{\partial^2
h}{\partial X^2}+\Gamma_y \frac{\partial^2 h}{\partial Y^2}, 
\end{equation}
where in our normalization $F$ is a proportionality constant
between $v_O$ and $\Gamma_0$, $\Gamma_x$, $\Gamma_y$, that can be found in 
the Appendix. Considering a periodic perturbation to the planar surface 
$h(X,Y,t\!=\!0)=A e^{i(k_1X+k_2Y)}$, and substituting this
expression into Eq.\ (\ref{varh}), the surface profile evolves as 
\begin{align}
&h(X,Y,t)=-\Gamma_0 \, t+Ae^{rt}e^{i(k_1X+k_2Y-\omega t)}, \\
&r=-\Gamma_xk_1^2-\Gamma_yk_2^2, \nonumber\\
&\omega=-\Gamma_0'k_1. \nonumber
\end{align}
If $\Gamma_x$ and/or $\Gamma_y$ are negative, there will be values for the 
wave-vector $(k_1,k_2)$ of the perturbation that make it grow exponentially. 
This behavior is a reflection of the well-known physical instability leading 
to ripple formation,\cite{sigmund:1973,bradley_harper:1988} due to the 
curvature dependence of the erosion velocity, that is larger in surface troughs
than in surface protrusions. The observed ripple wavelength arises when additional 
smoothing mechanims such as surface diffusion exist that compete with the sputter 
instability, leading to selection of a specific length-scale. Taking these 
mechanisms into account,\cite{bradley_harper:1988,makeev:2002} Eq.\ (\ref{varh}) reads
\begin{eqnarray}
\frac{\partial h}{\partial t} & \cong & 
F \, \left\{ -\Gamma_0(\theta)+\frac{\partial \Gamma_0(\theta)}{\partial
\theta} \frac{\partial h}{\partial X}+\Gamma_x \frac{\partial^2
h}{\partial X^2}+\Gamma_y \frac{\partial^2 h}{\partial Y^2} \right\} \nonumber \\
 & & -B \nabla^4 h, \label{varh+dif}
\end{eqnarray}
where, in principle, $B$ is a thermally activated coefficient which depends on
the surface self-diffusivity $D_s$, the free energy per unit area
$\gamma$ and the number of atoms per unit area moving across the
surface $\sigma$ as $B=2D_s\gamma\sigma/(n^2k_BT)$. In this case,
$r=-\Gamma_xk_1^2-\Gamma_yk_2^2-B(k_1^2+k_2^2)^2$, and there is
only a band of unstable perturbations. The observed ripple wavelength $\ell$ is 
provided by the wave-vector which has the largest {\em positive} value of $r$, 
and is proportional to 
$\sqrt{\frac{B}{F |\Gamma_x|}}$ or $\sqrt{\frac{B}{F |\Gamma_y|}}$ when 
$\Gamma_x<\Gamma_y<0$ ($\theta < \theta_c$) or $\Gamma_y<\Gamma_x$ 
($\theta > \theta_c$), respectively.

As we have seen in the previous section, the behavior of $\Gamma_x$ and $\Gamma_y$ is
similar for all the cases considered, while the qualitative behavior of the yield is 
quite different to that found by Bradley and Harper for amorphous or policrystalline 
substrates. Moreover, since dependences of the ripple wavelength $\ell \propto
\sqrt{\frac{B}{F |\Gamma|}}$ on parameters such as ion flux, $\Phi_0$, temperature,
or average ion energy, $\epsilon$, are due to those in the constants $F$ and $B$,
and these {\em are the same as those in BH} (see Appendix), Eq.\ (\ref{varh+dif})
predicts these for Cu to be (qualitatively) the same as obtained from BH 
theory.\cite{makeev:2002} Note this is {\em also} the case in the presence of 
non-thermal surface diffusion, in which, similarly to BH,\cite{makeev:2002} 
the constant $B$ has no dependence on temperature and is, rather, proportional
to $F$.

\section{Summary and Outlook}

We have studied numerically the sputtering process of Cu ions on Cu fcc crystals by
means of the binary collision approximation. We have analyzed the distribution of 
sputtered particles and their energies, and found significant deviations from Sigmund's 
formula, which is traditionally employed to study the sputtering process in the
framework of continuum theories, as applied to amorphous and policrystalline substrates.
In particular, we find that near the point where the ion penetrates the target, the 
sputter probability goes to zero, while the Bradley-Harper/Sigmund theory predicts 
maximum sputtering at that point.

We have fitted heuristic functions to our data. We find that an {\em exponential} 
(rather than {\em Gaussian} as in Sigmund's theory) decay with a combination of a 
quadratic and a linear prefactor fits the data well. The main physical effect, namely, 
the ``hole'' near the point of penetration, can be reproduced also qualitatively using a 
Gaussia distribution with with a quadratic prefactor, that is physically better defined 
than the heuristic fitting distribution, and lends itself to exact results. 
We have performed analytical calculations of the local erosion velocity following the 
Bradley-Harper approach for one- and two-dimensional surfaces, for both types of 
modified distributions (for the two-dimensional exponential distribution, the equation 
could be solved only numerically). We find that the sputter yield is
qualitatively different as compared the the BH approach. As a function of the angle
of incidence, the yield exhibits a maximum at an intermediate angle, and then decreases 
when approaching grazing incidence. This is in good agreement with experimental 
findings, in marked contrast with the analogous BH result using Sigmund's distribution, 
even without implementing explicitly reflection of the ions for grazing 
incidence, which is usually regarded as the main cause for the decay of the yield
at grazing incidence. Finally, we have computed also the ripple orientation-determining
parameters $\Gamma_x$, $\Gamma_y$, usually referred to in this context as 
effective surface tension parameters. These turn out to be only
slightly modified with respect to the BH theory, and lead to a qualitatively 
similar pattern formation process. Dependencies of the ripple wavelength on 
phenomenological parameters, such as ion flux, ion average energy,
and temperature are as in BH theory.\cite{makeev:2002} 
Since the influence of non-linearities on ripple characteristics is still under debate 
even within Sigmund's theory proper, we have not considered this type of effects here. 
At any rate, the same type of non-linear terms would appear in the interface Eq.\ 
(\ref{varh+dif}) as compared to the corresponding equation for amorphous or
policrystalline substrates.\cite{cuerno:1995,makeev:2002} 

Thus, as a general conclusion on pattern formation by ion-beam sputtering, our results 
justify the similarities found in experiments on metals, to the analogous processes in
amorphous or amorphizable materials, and point to potential quantitative differences
that would possibly merit further studies. Additional features of ripple formation
in metals, such as their existence for normal incidence or change of orientation 
with temperature \cite{rusponi:1997,rusponi:1998,rusponi:1998b} are {\em not} explained
by the special properties of the collision cascades in these systems that we have 
studied here but, rather, by the special properties of surface diffusion in such 
anisotropic substrates.

Regarding future work, 
it would be also interesting to see whether the hole near the 
point of penetration can be found in experiments, and/or in more detailed simulations
(such as e.g.\ by Molecular Dynamics). To our knowledge, no analysis of single-ion 
impacts on metals exist so far. Furthermore, it would be worth incorporating the 
modified energy distribution into existing simple Monte Carlo models of surface 
sputtering, such as those in Refs.\ \onlinecite{hartmann:2002}, 
\onlinecite{cuerno:1995b}, in order to improve their
description of erosion processes in metallic substrates, specially at the large
distance and long time regime for which this type of models is particularly suited.

\begin{acknowledgments}
This work has been supported by the Sonderforschungsbeeich 602 of the
Deutsche Forschungsgemeinschaft, and by MECD (Spain) grant No.\ BFM2003-07749-C05-01. 
A.\ K.\ H.\ obtained financial support from the {\em VolkswagenStiftung} (Germany) 
within  the program ``Nachwuchsgruppen an Universit\"aten''. A.\ K.\ H.\ 
thanks K.P. Lieb for helpful suggestions.
J.\ M.-G.\ 
acknowledges support from MECD (Spain) through an FPU fellowship.
\end{acknowledgments}

\appendix

\section{Analytic expressions for coefficients in the erosion velocity}

In this appendix, we provide the full expressions for the coefficients
appearing in various expressions for the surface erosion velocity, 
Eqs.\ (\ref{v1e}), (\ref{v1g}), and (\ref{vOg2d}), that have been 
computed analytically for those energy distributions for which such type of results 
are achievable.

\subsection{One-dimensional modified Gaussian distribution}

\label{apendiceA}
\begin{align}
    &v_O=N_g^{1d}\Lambda \epsilon \Phi_0\,e^{-\frac{a^2}{2\sigma_z^2}}
    \Big[\Gamma^{g,1d}_0 + \frac{\Gamma^{g,1d}_x}{R_x}\Big] \nonumber
\end{align}
\begin{align}
    &\Gamma^{g,1d}_0=\frac{\sqrt{\pi}e^{\frac{A^2_g}{4B_g}}(A^2_g+2B_g)}{4B_g^{5/2}}\cos^3\gamma_0,\nonumber
\end{align}
\begin{align}
    &\Gamma^{g,1d}_x=\frac{\sqrt{\pi}e^{\frac{A^2_g}{4B_g}}}{32B_g^{11/2}}\big[2A_g^3B_g(A_gb_g-10c_g)-A_g^5c_g-\nonumber\\
    &4A_gB_g^2(A^2a_g-6A_gb_g+15c_g)-24B_g^3(A_ga_g-b_g)\big],\nonumber
\end{align}
\begin{align}
    &A_g=\frac{a}{\sigma_z^2}\sin\gamma_0,\,B_g=\frac{1}{2\sigma_{z}^2}\sin^2\gamma_0+\frac{1}{2\sigma_x^2}\cos^2\gamma_0,\nonumber
\end{align}
\begin{align}
    &a_g=-2\sin\gamma_0 \cos^2\gamma_0,\,b_g=-\frac{a}{2\sigma_z^2}\cos^4\gamma_0,\nonumber
\end{align}
\begin{align}
    &c_g=\Big(\frac{1}{2\sigma_x^2}-\frac{1}{2\sigma_z^2}\Big)\cos^4\gamma_0\sin\gamma_0.\nonumber
\end{align}

\subsection{One-dimensional exponential distribution}
\label{apendiceB}
\begin{align}
    &v_O=N^{1d}_e\Lambda \epsilon \Phi_0\,e^{-\frac{a^2}{2\sigma_z^2}}
    \Big[\Gamma^{e,1d}_0 + \frac{\Gamma^{e,1d}_x}{R_x}\Big]\nonumber
\end{align}
\begin{align}
    &\Gamma^{e,1d}_0=\frac{a_e}{B_e}+\frac{1}{8B_e^{5/2}}\sum_{i=1,2}\Big\{-2\sqrt{B_e}A_{e,i}b_e+\nonumber\\
        &\big[{A^2_{e,i}}b_e-2a_eA_{e,i}B_e+2b_eB_e\big]\sqrt{\pi}e^{\frac{{A^2_{e,i}}}{4B_e}}\,{\rm erfc}\Big(\frac{A_{e,i}}{2\sqrt{B_e}}\Big)\Big\}\nonumber
\end{align}
\begin{align}
    &\Gamma^{e,1d}_x=\frac{1}{64B_e^{11/2}}\sum_{i=1,2}\Big\{2\sqrt{B_e}\big[(-1)^iA^4_{e,i}f_e-2A^3_{e,i}B_ee_{e,i}+\nonumber\\
        &+(-1)^i16B_e^3d_{e,i}-4B_e^2(A_{e,i}(5e_{e,i}+(-1)^i2B_ec_e))+\nonumber\\
        &(-1)^i18A^2_{e,i}B_ef_e+(-1)^i4A_{e,i}B_e^2d_{e,i}\big]+\nonumber\\
        &\big[-(-1)^iA^5_{e,i}f_e+2A^3_{e,i}B_e(A_{e,i}e_{e,i}-(-1)^i10f_e)+\nonumber\\
        &-4A_{e,i}B_e^2((-i)^iA^2_{e,i}d_{e,i}-6A_{e,i}e_{e,i}+(-1)^i15f_e)+\nonumber\\
        &8B_e^3((-1)^iA^2_{e,i}c_e-(-1)^i3A_{e,i}d_{e,i}+3e_{e,i})\big]
\times\nonumber\\
&\sqrt{\pi}e^{\frac{A^2_{e,i}}{4B_e}}{\rm erfc}
\Big(\frac{{A_{e,i}}}{2\sqrt{B_e}}\Big)\Big\}\nonumber
\end{align}
\begin{align}
    &A_{e,1}=\frac{\cos\gamma_0}{\sigma_x}-\frac{a\sin\gamma_0}{\sigma_z^2},\,A_{e,2}=\frac{\cos\gamma_0}{\sigma_x}+\frac{a\sin\gamma_0}{\sigma_z^2}\nonumber
\end{align}
\begin{align}
    &B_e=\frac{\sin^2\gamma_0}{2\sigma_s^2},\, a_e=c\cos^2\gamma_0,\, b_e=\cos^3\gamma_0.\nonumber
\end{align}
\begin{align}
    &c_e=-\frac{3}{2}c\cos(\gamma_0)\sin(\gamma_0),\,\nonumber
\end{align}
\begin{align}
    &d_{e,1}=\big(\frac{c}{2\sigma_x}-2\big)\cos^2(\gamma_0)\sin(\gamma_0)+\frac{c\,a}{2\sigma_z^2}\cos^3(\gamma_0),\nonumber
\end{align}
\begin{align}
    &d_{e,2}=\big(\frac{c}{2\sigma_x}-2\big)\cos^2(\gamma_0)\sin(\gamma_0)-\frac{c\,a}{2\sigma_z^2}\cos^3(\gamma_0),\nonumber
\end{align}
\begin{align}
    &e_{e,1}=\big(\frac{c}{2\sigma_z^2}-\frac{1}{2\sigma_x}\big)\cos^3(\gamma_0)\sin(\gamma_0)-\frac{a}{2\sigma_z^2}\cos^4(\gamma_0),\nonumber
\end{align}
\begin{align}
    &e_{e,2}=-\big(\frac{c}{2\sigma_z^2}+\frac{1}{2\sigma_x}\big)\cos^3(\gamma_0)\sin(\gamma_0)-\frac{a}{2\sigma_z^2}\cos^4(\gamma_0),\nonumber
\end{align}
\begin{align}
    &f_e=2\cos^5(\gamma_0)\sin(\gamma_0).\nonumber
\end{align}

\subsection{Two-dimensional modified Gaussian distribution}
\label{apendiceC}
\begin{align} \label{DvOg2d}
    v_O&=N_g\Lambda \epsilon \Phi_0\,e^{-\frac{a^2}{2\sigma_z^2}}
    \Big[\Gamma^g_0 +
    \frac{\Gamma^g_x}{R_x}+\frac{\Gamma^g_y}{R_y}\Big]\nonumber
\end{align}
\begin{align}
    \Gamma^g_0&=\frac{\pi e^{\frac{A^2_g}{4B_g}}
(b_0A^2_g+4a_0B_g^2+2b_0B_g)}{2\sqrt{2}B_g^{5/2}},\nonumber
\end{align}
\begin{align}
     \Gamma^g_x&=\frac{\pi\,e^{\frac{A^2_g}{4B_g}}}{16\sqrt{2}B_g^{11/2}}\Big\{2B_g\big[4B_g^2b_x(A_g^2+2B_g)-8A_gB_g^3a_x+\nonumber\\
        &-2A_gB_gc_x(A_g^2+6B_g)+d_x(A_g^4+12A_g^2B_g+12B_g^2)\big]+\nonumber\\
        &-A_ge_x(A_g^4+20A_g^2B_g+60B_g^2)\Big\}\nonumber
\end{align}
\begin{align}
    \Gamma^g_y&=\frac{\pi\,e^{\frac{A^2_g}{4B_g}}}{4\sqrt{2}B_g^{7/2}}\big[2B_gc_y(A_g^2+2B_g)-4B_g^2a_y(b_y-2B_g)+\nonumber\\
        &-A_gd_y(A_g^2+6B_g)\big]\nonumber
\end{align}
\begin{align}
    A_g&=\frac{a}{\sigma_z^2}\sin\gamma_0,\,B_g=\frac{1}{2\sigma_{z}^2}\sin^2\gamma_0+\frac{1}{2\sigma_{xy}^2}\cos^2\gamma_0,\nonumber
\end{align}
\begin{align}
    a_0&=\sigma_{xy}^3\cos\gamma_0,\,b_0=\sigma_{xy}\cos^3\gamma_0, \nonumber
\end{align}
\begin{align}
    a_x&=-\sigma_{xy}^3\sin\gamma_0,\, b_x=-\frac{\sigma_{xy}^3}{2\sigma_z^2}a\cos^2\gamma_0,\nonumber
\end{align}
\begin{align}
    c_x&=-\Big(\frac{3}{2}\sigma_{xy}+\frac{\sigma_{xy}^3}{2\sigma_z^2}\Big)\cos^2\gamma_0\sin\gamma_0\nonumber
\end{align}
\begin{align}
    d_x&=-\frac{\sigma_{xy}}{2\sigma_z^2}a\cos^4\gamma_0,\,e_x=\Big(\frac{1}{2\sigma_{xy}}-\frac{\sigma_{xy}}{2\sigma_z^2}\Big)\sin\gamma_0\cos^4\gamma_0\nonumber
\end{align}
\begin{align}
    a_y&=-\frac{3\sigma_{xy}^5}{2\sigma_z^2}a\cos^2\gamma_0\nonumber
\end{align}
\begin{align}
    b_y&=-\frac{3\sigma_{xy}^5}{2\sigma_z^2}\cos^2\gamma_0\sin\gamma_0+\frac{\sigma_{xy}^3}{2}\cos^2\gamma_0\sin\gamma_0\nonumber
\end{align}
\begin{align}
    c_y&=-\frac{\sigma_{xy}^3}{2\sigma_z^2}a\cos^4\gamma_0\nonumber
\end{align}
\begin{align}
    d_y=\Big(\frac{\sigma_{xy}}{2}-\frac{\sigma_{xy}^3}{2\sigma_z^2}\Big)\cos^4\gamma_0\sin\gamma_0. \nonumber
   \end{align}

\bibliography{main2.2}

\begin{thebibliography}{35}
\bibitem[{foo()}]{footnote1}
\bibinfo{note}{We performed also some test simulations for Ar ions and obtained
  qualitatively the same results. Hence, we show only the results for Cu ions
  here.}
\end{thebibliography}

\end{document}